\documentclass[twocolumn, secnumarabic, amssymb, noshowpacs, aps, showpacs,prb]{revtex4-1}
\usepackage{graphics}
\usepackage[dvips]{graphicx}
\usepackage{dcolumn}
\usepackage{amsmath}
\usepackage{lipsum}

\begin{document}

\title{Photonic transmittance in metallic and metamaterial Superlattices}
\author{Pedro Pereyra}
\address{F\'{i}sica Te\'{o}rica y Materia Condensada, UAM-Azcapotzalco, Av. S. Pablo 180, C.P. 02200, M\'{e}xico D. F., M\'{e}xico }
\date{\today}

\begin{abstract}

Recent developments in optical devices, photonic crystals and thin-film epitaxial-growing techniques, have spurred
theoretical and experimental research on the subject of electronic and electromagnetic wave transport through
multilayered structures containing dielectric, semiconductor, left-handed, metallic (lossy) media and metamaterials.
We present here the transmission of electromagnetic waves through layered structures of metallic and left-handed media. Based on the theory of finite periodic systems, we show that besides the strong influence of the incidence angle, the low transmission characteristic of a single conductor slab, for frequencies $\omega$ below the plasma frequency $\omega_p$, becomes in this domain highly oscillating and eventually transparent when the photonic superlattice parameters match certain conditions. Similarly, the well established transmission coefficient  of a single left-handed slab, that exhibits optical antimatter effects, becomes highly resonant with superluminal effects in superlattices with more than one unit cell. We determine the space-time evolution of a wave packet through the $\lambda/4$ photonic superlattice whose transmission coefficent is a sequence of isolated and equidistant peaks with negative phase times. We show that the space-time evolution of a Gaussian wave packet, with centroid at any of these peaks, agrees with the theoretical predictions and no violation of the causality principle occurs.

We study the transmission of electromagnetic waves through layered structures of metallic and left-handed media. Resonant band structures of transmission coefficients are obtained as functions of the incidence angle, the geometric parameters, and the number of unit cells of the superlattices. The theory of finite periodic systems that we use is free of assumptions, the finiteness of the periodic system being an essential condition. We rederive the correct recurrence relation of the Chebyshev polynomials that carry the physical information of the coherent coupling of plasmon modes and interface plasmons and surface plasmons, responsible for the photonic bands and the resonant structure of the surface plasmon polaritons. Unlike the dispersion relations of infinite periodic systems, which at best predict the bandwidths, we show that the dispersion relation of this theory predicts not only the bands, but also the resonant plasmons' frequencies, above and below the plasma frequency. We show that besides the strong influence of the incidence angle and the characteristic low transmission of a single conductor slab, for frequencies $\omega$ below the plasma frequency $\omega_p$, we find that in the low frequencies domain, the coherent coupling of the bulk plasmon modes and the interface surface plasmon polaritons lead to oscillating transmission coefficients, and depending on the parity of the number of unit cells $n$ of the superlattice, the transmission vanishes or amplifies as the conductor width increases. Similarly, the well-established transmission coefficient of a single left-handed slab, which exhibits optical antimatter effects, becomes highly resonant with superluminal effects in superlattices with more than one unit cell. We determine the space-time evolution of a wave packet through the $\lambda/4$ photonic superlattice whose bandwidth becomes negligible, and the transmission coefficient becomes a sequence of isolated and equidistant peaks with negative phase times. We show that the space-time evolution of a Gaussian wave packet, with the centroid at any of these peaks, agrees with the theoretical predictions, and no violation of the causality principle occurs. \end{abstract}


\maketitle

\section{Introduction}
For many years, photonic crystals (PC) of metal-dielectric structures designed to control and manipulate the propagation of electromagnetic fields have been widely studied, both theoretically and experimentally\cite{Yablonovitch1987,Pendry1994,Baba1999,Gadner1999,BottenI2000,
BottenII2000,Joanopoulos1995,Kawakami2003,Jia-Yasumoto2005}. Among the various methods, used to calculate the transmittance or reflectance of photonic crystals, the scattering matrix method based on multipole expansions stands out, which has been further developed and extended according to the particular geometries of the PC structures. The properties and physical meaning of the scattering matrix agree with the geometries of the devices to which it could be applied, i.e., to photonic crystals whose structures are circular cross-sections (in 2D photonic crystals) or crystals formed by spherical inclusions. The bulk of the physical results reported within these approaches are mostly numerical, providing little insight into the underlying physics of the electromagnetic (EM) wave propagation. Many papers were published on metallic superlattices \cite{Camley1984,Vigneron1985,Xue1985,Wallis1987,
Nazarov1994,Quinn1995,Bria2004}. A common feature of these papers is that they end up dealing with infinite or semi-infinite superlattices by introducing the approximate Bloch periodicity condition, whose first drawback is the derivation of dispersion relations that give rise to {continuous} sub-bands, i.e., to Kronig--Penney-like dispersion relations that give at best the widths of the allowed and forbidden sub-bands. Although the infinite periodic system approximation could be justified for macroscopic (bulk) crystals, where the number of unit cells is truly large, it is hard to justify for SLs
 where the number of unit cells is of the order of a dozen. Strictly speaking, even basic quantities like reflection and transmission coefficients are impossible to conceive of for infinite systems. A well-established method to study the transmission of electromagnetic waves through layered and periodic systems is the transfer matrix method \cite{Abeles1948,Yeh1988}. Different versions for different kinds of applications of this approach have been implemented. For photonic crystals, we find, among others, the rather cumbersome transfer matrix method introduced by Pendry for cylindrical dielectric arrays and the transfer matrices defined in terms of reflection and transmission amplitudes introduced by Botten et al. Others, like those in \cite{Wendler1987,Mochan1988,
Trutschel1989,Sheng1992}, start well, obtaining the unit-cell transfer matrices, but when they have to deal with a superlattice, their theoretical approach becomes extremely cumbersome or decide to follow P.
 Yeh's flawed argument \cite{Yeh1988} reintroducing the unnecessary Floquet theorem, the Bloch functions, and the Kronig--Penney-like dispersion relation in the transfer matrix approach.

In the theory of finite periodic systems (TFPS), the finiteness property of the actual periodic system is an essential condition. In open systems, the resonant transmission coefficients have a simple and neat relation with the resonant dispersion relation, while in bounded systems, the energy eigenvalues' equations predict not only the sub-bands, but also the whole structure of intra-sub-band and surface energy levels \cite{Pereyra2005}. In this theory, one can also determine, analytically and without any approximation, the corresponding resonant functions, the eigenfunctions, surface states, and a number of closed formulas for an accurate calculation of the SL transmission coefficients. This theory, applied to metallic superlattices and left-handed media superlattices, allows us to determine the intra-sub-band plasma modes, the localized and resonant surface plasmons, as well as the coherent coupling of the plasma modes that give rise to interesting features of the band structures for the transmittance of EM waves and tunneling times as functions of the SL parameters, the incidence angle, and the number of unit cells. It is worth noting that all the closed expressions derived in the TFPS and used to study the space-time evolution of EM fields through layered metal-dielectric and left-handed-dielectric structures are exact, i.e., no approximation is required once the unit cell transfer matrix is given. We will show that the metal-dielectric planar photonic superlattices exhibit, basically, the same properties of the complex photonic crystals.

In recent years, the space-time evolution of electromagnetic waves through layered structures and superlattices of right- and left-handed media (LHM and RHM) has been studied in the framework of the theory of finite periodic systems (TFPS) \cite{Pereyra2008, Simanjuntak2007}. The transmittance through layered structures containing metallic slabs is rather complicated because of the complex indices, the incidence angle, the cutoff frequency, etc. In \cite{Pereyra2008}, we gave a brief introduction to the propagation of EM waves through this type of system. Here, we will extend the analysis and present results for the transmission of EM waves through $(air/metal/air
)^n$ superlattices. Applying the theory of finite periodic systems, we will obtain the transmission and reflection coefficients as functions of the various parameters of the photonic SLs. The characteristic stopbands and resonant transmission as a function of the EM wave frequency, the incidence angle, the slabs thicknesses, etc., will be found.

An important quantity, with clear consequences in the transport properties of layered structures, is the transmission amplitude phase $\theta_t$. It has been shown that the superlattice phase times $\tau$, defined as the frequency derivative of $\theta_t$ (see \cite{Pereyra2000}), account, within the experimental error of $\sim 0.1 $fs, for the observed tunneling times \cite{Spielmann1994}. In \cite{Pereyra2011}, explicit realizations of the optical antimatter behavior upheld by Pendry and Ramakrisnan \cite{Pendry2003} were observed. Performing a kind of ``theoretical experiment'', the antimatter behavior was shown for a wave packet moving through a sequence of two slabs of equal thickness and opposite refraction indices, placed adjacent to one another.

In this paper, we will also discuss the transport and transmission time of Gaussian electromagnetic wave packets through an $air
(LR)^n air
$ structure. This issue was partially considered in \cite{Pereyra2019}. In this structure, $(LR)^n$ refers to a superlattice of left- and right-handed media, with refractive indices and widths $n_L=-|n_L|$, $n_2$, $d_L$, and $d_2$, respectively. We will show that when the superlattice parameters, fulfill the $\lambda/4$ condition, the band structure of the transmission coefficient becomes a sequence of isolated and equidistant resonances (IER), with negative tunneling times (NTT) and narrow, practically, vanishing bands. Following the actual evolution of a wave packet through an SL with NTT, we show that no violation of the causality principle occurs.

In Section \ref{section2}, we refer to the transfer matrix for an EM wave in parallel polarization through a conductor slab bounded by dielectric media, and we recall the relevant formulas of the TFPS. In Section \ref{section3}, we calculate the transmission and reflection coefficients for the metal-dielectric superlattice. In Section \ref{section4}, we outline the scattering amplitudes for superlattices containing right- and left-handed media slabs alternating with dielectrics and discuss the transport of Gaussian the EM wave packets through an $air(LR)^n air$ structure, or metamaterial superlattice (MMSL). For the benefit of the reader, we will repeat the neat derivation of the correct Chebyshev polynomials' recurrence relation. In Appendix A, we show that the zeros of the Chebyshev polynomials determine the resonant plasmons and a dispersion relation that gives not only the widths of the sub-bands, but also the intra-sub-band resonant states. Appendix C
 shows how the sub-bands and intra-sub-band resonances behave in the $\lambda /4$ limit.


\section{Electromagnetic Wave through $(air/metal/air)^n$ Structures}\label{section2}

\begin{figure*}
\begin{center}
\includegraphics[width=32pc]{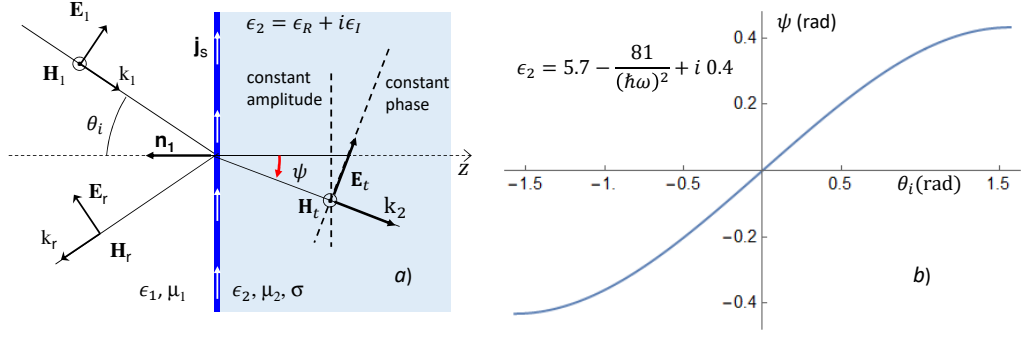}
\caption{Incoming, reflected, and transmitted fields at the interface dielectric-conductor. b) The incidence $\theta_i$ and the effective refraction angle $\psi$ for a silver slab with the dielectric constant shown here. }\label{fig1}
\end{center}
\end{figure*}

In this section, we discuss some interesting properties concerning the optical transmission of electromagnetic
waves through complex index media, for an EM field in parallel polarization, as shown in Figure \ref{fig1}.

If $z=0$ is the interface between a dielectric medium, say air, and a conductor with dielectric constant $\epsilon_2=\epsilon_R+i \epsilon_I$, magnetic permeability $\mu_2$, and conductivity $\sigma$, an EM wave, coming from $z<0$ with an incidence angle $\theta_i$,
moves in the conducting medium as shown in Figure \ref{fig1} a), where the constant amplitude planes are parallel to the reflecting surface, and the constant
phase planes defined by an effective refraction angle $\psi$ are given by \cite{Stratton1941}:
\begin{equation}
\tan \psi=\frac{k_1\sin \theta_i}{q}.
\end{equation}
\begin{widetext}
Here, $k_1=\omega \sqrt{\epsilon_1\mu_1}$ is the wave vector of the incident EM field, $q=\rho (\epsilon_R \cos \gamma-\epsilon_I \sin \gamma)$ with:
\begin{equation}
\rho=\Bigl[1+\left(\frac{k_1^2}{k_2^2}\sin^2\theta_i\right)^2-
2\frac{k_1^2}{k_2^2}\sin^2\theta_i \cos 2\theta_{2}\Bigr]^{1/4}, \hspace{0.5in}
\gamma=\frac{1}{2}\tan^{-1}\Bigl[\frac{k_1^2\sin^2\theta_i \sin 2\theta_{2}}{k_2^2-k_1^2\sin^2\theta_i \cos 2\theta_{2}}\Bigr],
\end{equation}
\end{widetext}
and:
\begin{equation}
k_2=\pm \omega\left[\mu_2\sqrt{\epsilon_R^2+\frac{\sigma^2}{\omega^2}}\, \right]^{1/2}
\hspace{0.5in}\theta_{2}=\frac{1}{2}\tan^{-1}\Bigl[\frac{\sigma}{\omega \epsilon_R}\Bigr].
\end{equation}

In Figure \ref{fig1} b), the effective refraction angle $\psi$ is shown as a function of the incoming angle $\theta_i$. The behavior of this angle and the parameters that rise when the EM wave enters a conductor slab with a complex dielectric function are well known \cite{Stratton1941}. In order to study the transport of the electromagnetic waves through the finite superlattices, we use the theory of finite periodic systems. In this theory, the whole superlattice transfer matrix $M_n$ and the various closed formulas for the relevant physical quantities require the transfer matrix of a unit cell of the SL. If the conducting slab has a thickness $d_c$, the transfer matrix that connects the EM field at the left with the EM field at the right is:
\begin{widetext}

\begin{equation}\tag{4}
M_c=\frac{1}{2 k_1\mu_2 \cos\psi+2\xi}\left(\begin{array}{cc} \alpha_l & \beta_l \cr  \beta_l^*  & \alpha_l^* \end{array}\right)\left(\begin{array}{cc} e^{i( q+i p) d_c} & 0 \cr 0 & e^{-i( q -i p)d_c} \end{array}\right)\frac{1}{2 \kappa\mu_1 \cos\theta_i}\left( \begin{array}{cc} \alpha_l^* & -\beta_l \cr -\beta_l^* & \alpha_l \end{array}\right),
\end{equation}
where  $p=\rho (\epsilon_R \sin \gamma+\epsilon_I \cos \gamma)$, $\kappa=\bigl(q^2+k_1^2\sin^2\theta_i \bigr)^{1/2}$, $\xi=k_1\mu_2\sec\theta_i \tan\psi$ and
\begin{equation}\tag{5}
\begin{aligned}
 \alpha_l&=&  k_1\mu_2 \sec\theta_i +\kappa \mu_1 \cos\psi+ \xi+i\, p\mu_1 \cr
 \beta_l &=& k_1\mu_2 \sec\theta_i-\kappa \mu_1 \cos\psi-\xi+i\, p\mu_1 .
 \end{aligned}
\end{equation}
\end{widetext}
with slab transfer matrix $M_c$
\begin{equation}\tag{6}
M_c=\left( \begin{array}{cc} \alpha_c & \beta_c \cr \beta_c^* & \alpha_c^* \end{array}\right). \nonumber
\end{equation}
and unit-cell transfer matrix
\begin{equation}\tag{8}
M=\left( \begin{array}{cc} \alpha & \beta \cr \beta^* & \alpha^* \end{array}\right)=\left( \begin{array}{cc} e^{i k_1d_a\cos \theta_i}\alpha_c & \beta_c \cr \beta_c^* & e^{-i k_1d_a\cos \theta_i}\alpha_c^* \end{array}\right).
\end{equation}

It is well known from the transfer matrix approach that the transmission and reflection coefficients through a single conductor slab are given by:
\begin{equation}
T_c=\frac{1}{|\alpha_c|^2}\hspace{0.3in}{\rm{and}}\hspace{0.3in}
R_c=\frac{|\beta_c|^2}{|\alpha_c|^2}.
\end{equation}

Given the transfer matrix $M_c$, it is easy to obtain the transfer matrix of a unit cell of a superlattice. In fact, if a unit cell is a conductor slab bounded by equal dielectric layers of thickness $d_a/2$, the transfer matrix of the unit cell is:
\begin{equation}
M=\left( \begin{array}{cc} \alpha & \beta \cr \beta & \alpha^* \end{array}\right)=\left( \begin{array}{cc} e^{i k_1d_a\cos \theta_i}\alpha_c & \beta_c \cr \beta_c & e^{-i k_1d_a\cos \theta_i}\alpha_c \end{array}\right).
\end{equation}

At this point, it is worth making clear that in the transfer matrix approach, we can distinguish, at least, two alternatives. One is to follow a faulty argument in P. Yeh's book, which reintroduces the Floquet theorem in exchange for the Kramer condition $|Tr M|\leq 2$, which gives rise to a Kronig--Penney-like dispersion relation. In this alternative, one obtains, at most, the bandwidths of continuous sub-bands for infinite superlattices. The other alternative is to follow the TFPS and derive analytically the correct Chebyshev polynomial recurrence relation, in terms of which one writes the transfer matrix for the SL with $n$ unit cells and the physical quantities. Since the resilience of the standard approach followers is strong, let us, for the benefit of the reader, repeat here the derivation of the Chebyshev polynomials' recurrence relation, which was long ago \cite{Pereyra1998,Pereyra2002,Pereyra2012} published for an arbitrary number of propagating modes. Suppose that:
\begin{eqnarray}
M =
 \left(
\begin{array}{ccll}
\alpha& \beta \\
 \beta^{\ast}& \alpha^{\ast}
\end{array}\right),
\end{eqnarray}
is the transfer matrix of a unit cell.\footnote{The symmetry of this matrix corresponds to time reversal invariant systems. The proof in other symmetries is similar.}
 The transfer matrix of a sequence of $n$ unit cells is:
\begin{eqnarray}
 M_{n} = \underbrace{M.M...M}_{ n \rm\; factors} = M^{n}.
\end{eqnarray}

This product of matrices can be written in different ways, for example we can write $M_n$ as the product of $M_{n-1}$ with the matrix $M$, i.e.:
\begin{eqnarray}
 M_{n} = \underbrace{M.M...M}_{ n\rm-1 \; factors}M = M_{n - 1}M.
\end{eqnarray}

In terms of the matrix elements, this product is:
\begin{eqnarray}
 \left(
\begin{array}{ccll}
\alpha_{n}& \beta_{n} \\
 \beta^{\ast}_{n} & \alpha^{\ast}_{n}
\end{array}\right) = \left(
\begin{array}{ccll}
\alpha_{n - 1}& \beta_{n - 1} \\
 \beta^{\ast}_{n - 1} & \alpha^{\ast}_{n - 1}
\end{array}\right) \left(
\begin{array}{ccll}
\alpha& \beta \\
 \beta^{\ast} & \alpha^{\ast}
\end{array}\right).\label{Mnm1M}
\end{eqnarray}

Our purpose is to obtain $\alpha_{n}$ and $\beta_{n}$, provided that $\alpha$ and $\beta$ are known. From this equation, we have \footnote{For systems with more than one propagating mode, $\alpha$ and $\beta$ are matrices. More general cases were considered in \cite{Pereyra1998,Pereyra2002,Pereyra2012}.}:
\begin{eqnarray}\label{ecprec}
\alpha_{n} = \alpha_{n - 1}\alpha + \beta_{n - 1}\beta^{\ast},
\end{eqnarray}
and
\begin{eqnarray}
\beta_{n} = \alpha_{n - 1}\beta + \beta_{n - 1}\alpha^{\ast}.
\end{eqnarray}

If in the last equation, we solve for $\alpha_{n - 1}$, and we have:
\begin{eqnarray}\label{alfas}
\alpha_{n - 1} = \beta^{-1}\beta_{n} - \alpha^{\ast}\beta^{-1}\beta_{n - 1} 
\end{eqnarray}
thus 
\begin{eqnarray}
\alpha_{n } = \beta^{-1}\beta_{n+1} - \alpha^{\ast}\beta^{-1}\beta_{n}.\label{ecalphanm1}
\end{eqnarray}

Replacing these ${\alpha_n}$'s in the first equation of (\ref{ecprec}) and taking into account that $\alpha \alpha^*-\beta \beta^*=1$, we end up with the interesting three terms' recurrence relation:
\begin{eqnarray}
\beta^{-1}\beta_{n+1} - \left(\alpha + \alpha_{\ast}\right)\beta^{-1}\beta_{n} + \beta^{-1}\beta_{n-1} = 0.
\end{eqnarray}

Since $M^0=I$, $M^1=M$, and $\alpha + \alpha^{\ast} = 2\alpha_{\rm R}$, the last equation is nothing else but the recurrence relation of the Chebyshev polynomial of the second kind, evaluated at the real part of $\alpha$. In fact, if:
\begin{eqnarray}\label{betn}
\beta_{n}=\beta U_{n-1},
\end{eqnarray}
Equation (\ref{alfas}) becomes:
\begin{eqnarray}\label{alfn}
\alpha_{n } = U_{n} - \alpha^{\ast}U_{n-1},
\end{eqnarray}
and Equation (7) can be written in the most familiar notation:
\begin{eqnarray}
U_{n} - 2\alpha_R U_{n-1} + U_{n-2} = 0.
\end{eqnarray}
with the initial conditions $U_{0} = 1$ and $U_{-1} = 0$. It is important to notice that no approximation was introduced. There is no need to introduce quantities that are valid only for infinite systems, like the Floquet or Bloch theorem. The finiteness of the system is present through the order of the Chebyshev polynomials. Notice also that to
determine these polynomials and, consequently, the transfer matrix of the whole $n$ cell system (see Eqs. (10)), {it is enough to know the transfer matrix of the unit cell}. These results can be applied for systems with any number of unit cells and any potential profile (or refractive indices), and as will be seen below and was shown in \cite{Pereyra2005}, they allow obtaining accurate values for the resonant energies and wave functions (for open SLs) and accurate energy eigenvalues and eigenfunctions for bounded SLs. Given the matrix elements $\alpha_n$ and $\beta_n$ from Equations (\ref{betn}) and (\ref{alfn}), the transfer matrix of a (time reversal invariant) superlattice with $n$ unit cells is:
\begin{eqnarray}
\!M_n\!=\!\left(\begin{array}{cc} \alpha_n & \beta_n \cr \beta_n^* & \alpha_n^*\end{array}\right)=\left( \begin{array}{cc} U_n-\alpha^*U_{n-1} & \beta U_{n-1} \cr \beta^* U_{n-1} & U_n-\alpha U_{n-1} \end{array}\right).\nonumber \\
\end{eqnarray}

The transmission and reflection coefficients are:
\begin{equation}
T_n=\frac{1}{|\alpha_n|^2}=\frac{1}{|U_n-\alpha^*U_{n-1}|^2}
\end{equation}
\begin{equation}
R_n=\frac{|\beta_n|^2}{|\alpha_n|^2}=\frac{|\beta U_{n-1}|^2}{|U_n-\alpha^*U_{n-1}|^2},
\end{equation}

It is worth noticing that the $n$ cells' transfer matrix in Yeh's book\cite{Yeh1988} is formally similar, but with Chebyshev polynomials defined in terms of $K\Lambda$, the Bloch wavenumber $K$ and the SL periodicity $\Lambda$. His transfer-matrix approach describes infinite superlattices with dispersion relations that predict at best the widths of continuous bands.
\begin{widetext}
In Appendix A, we show that the transmission and reflection coefficients can be written also as:
\begin{equation}\label{ChebRes}
T_n=\frac{1}{1+|\beta|^2|U_{n-1}(\alpha_R)|^2}\hspace{0.3in}{\rm{and}}\hspace{0.3in}
R_n=\frac{|\beta|| U_{n-1}(\alpha_R)|^2}{1+|\beta|^2|U_{n-1}(\alpha_R)|^2},
\end{equation}

Here, it is easy to see that wherever $\alpha_R$ is a zero of the Chebyshev polynomial $U_{n-1}$, we have a resonance with $T_n=1$ and $R_n=0$. This property leads also to the resonant dispersion relation:
\begin{equation}
\cos \frac{\nu+(\mu-1)n}{n}\pi=(\alpha_R)_{\mu,\nu}\hspace{0.3in} {\rm with} \hspace{0.3in}\mu=1,2,3,... \;\; \nu=1,2,...,n-1.
\end{equation}

The quantum numbers $\mu$ and $\nu$ define the resonant energies $E_{\mu,\nu}=\hbar \omega_{\mu,\nu}$, of the $\nu^{\text{th}}$ intra-sub-band energy level of the sub-band $\mu$. Generally, $\mu=$1, 2, 3, ... and $\nu=$1, 2, ..., $n$ 1. Solving the dispersion relation, one easily obtains not only the bandwidths, but also all the resonant frequencies, or resonant energies \cite{Pereyra2005,Pereyra2002,Pereyra2017}.

In the particular case of the metallic superlattice that we are studying here, the resonant dispersion relation is:

\begin{eqnarray}\label{RDR}
\cos \frac{\nu \!+\!(\mu\!-\!1)n}{n}\pi\!=\!e^{\!-d_c p}\left[\cos d_c q\cos (d_ak_1\cos \theta_i)\!-\!\frac{\sin d_c q\sin (d_ak_1\cos \theta_i)}{4k_1\kappa\mu_1\mu_2}\left( \kappa^2 \mu_1^2\frac{\cos \theta_i}{\cos \psi}+k_1^2 \mu_2^2 \frac{\cos \psi}{\cos \theta_i}\right) \right]\cr
\end{eqnarray}
\end{widetext}

\begin{figure*}
\begin{center}
\includegraphics[width=36pc]{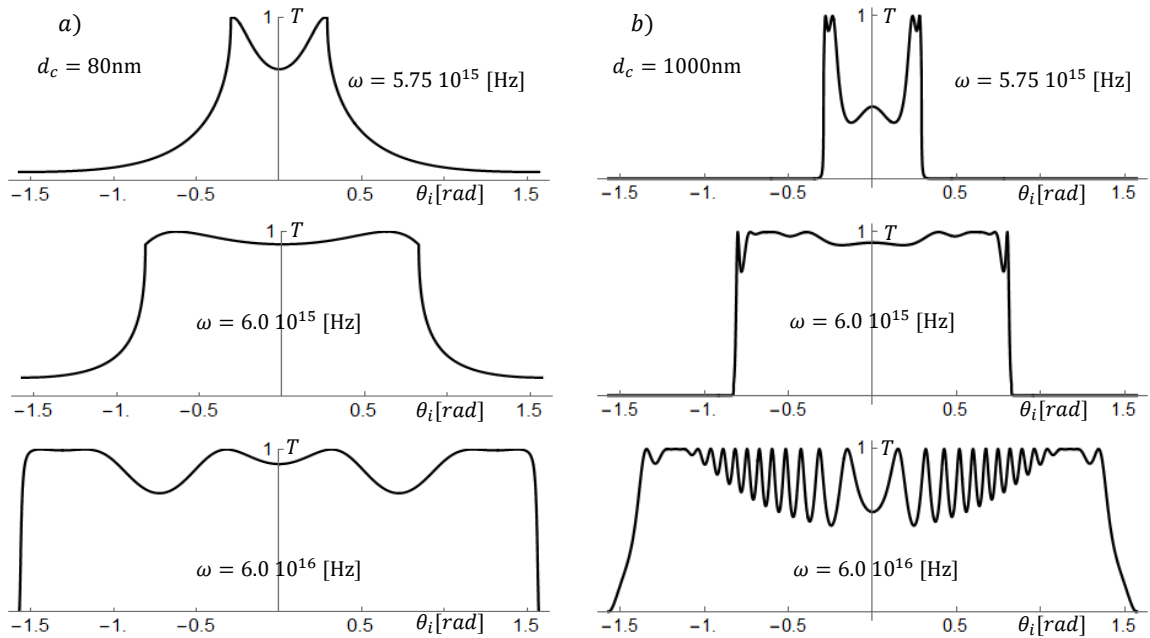}
\caption{Transmittance as a function of the frequency through a single slab of silver with thickness $d_c=$30nm (left-hand side graph) and $d_c=$80nm (right-hand side graph), for incidence angles $\theta_i=\pi/6,$ $\pi/4$, $\pi/3$, and slightly less than $\pi/2$. For the slab thicknesses considered here, the EM field is highly attenuated below the plasma frequency $\omega_p=$ 5.75 10$^{15}$rad/s and oscillating for $\omega > \omega_p$. Near the incidence angle of $\pi/2$, the narrow and isolated resonances correspond to localized surface plasmons. Notice that as the incidence angle grows, the resonances move towards the isolated surface plasmon resonances.}\label{fig2}
\end{center}
\end{figure*}

In the next section, we present some specific results for these quantities.

\section{Photonic Transmittance through Metallic Superlattices}\label{section3}

\begin{figure*}
\begin{center}
\includegraphics[width=36pc]{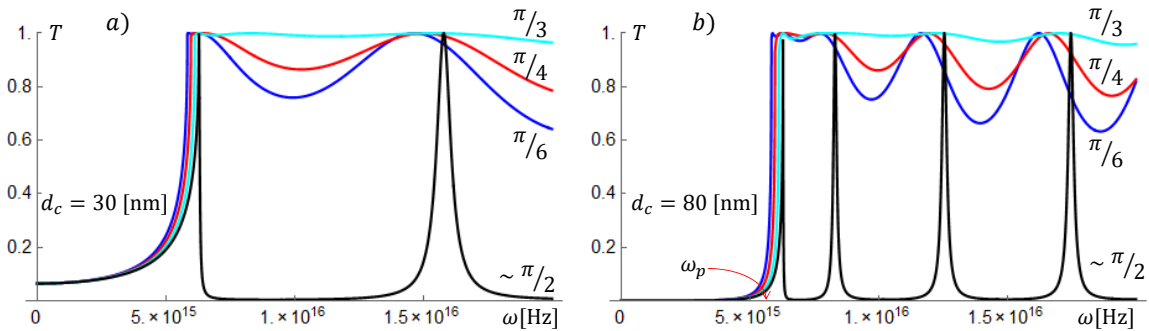}
\caption{Transmittance through a single slab as a function of the incidence angle $\theta_i$, for three different EM frequencies. For the graphs in the left-hand side column, the silver slab thickness is $d_c=$80nm, and in the right-hand side column, $d_c=$1000nm. Notice that the number of oscillations at high frequencies is of the order of $d_c/\lambda$. }\label{fig3}
\end{center}
\end{figure*}

In order to calculate the transmittance through metallic superlattices, we will assume that the metallic layers in the superlattice are thin silver films separated by also thin air films. The dielectric function that we use for silver is:
\begin{equation}
\epsilon_2=5.7-\frac{81}{F^2}+i \frac{\sigma}{\omega},
\end{equation}
the real part of which was taken from \cite{Pendry2000}. Here, $F=\hbar \omega$ is the frequency of the incoming wave expressed in units of eV. As shown in Figure \ref{fig3}, the transmission is null below the cutoff frequency ($5.75 10^{15}$ rad/s) where all incident radiation is reflected.

We will start calculating the transmittance of EM fields through a single silver slab, and we will calculate the transmittance through SLs with a finite number of unit cells. Our results are shown for a wider spectra of superlattice and EM field frequencies. We will consider EM fields with frequencies varying from ultra small to THz and conductor and air slabs with thicknesses $d_c$ and $d_a$, respectively, varying between 10nm to 1000nm.
In Figure \ref{fig2} a) and b), we have the transmittance through single silver slabs as functions of the frequency and for incidence angles $\theta_i=\pi/6,$ $\pi/4$, $\pi/3$, and slightly less than $\pi/2$. For the graphs in Figure \ref{fig2} a), the slab width is 30nm, while for Figure \ref{fig2} b), the slab width is 80nm. The EM field is highly attenuated for frequencies below the plasma frequency $\omega_p=$ 5.75 10$^{15}$rad/s. For frequencies above $\omega_p$, the transmittance is resonant, with values close to one. For $\theta_i=\pi/3$, the slab is almost transparent. This is compatible with the resonant transmissions observed in Figure \ref{fig3}, for $\omega$=6 10$^{15}$rad/s. We see also that for $\theta_i\simeq \pi/2$, the transmittance is characterized by narrow and non-overlapping resonances. The narrow peaks correspond to highly localized surface plasmon modes.
\begin{figure*}
\begin{center}
\includegraphics[width=36pc]{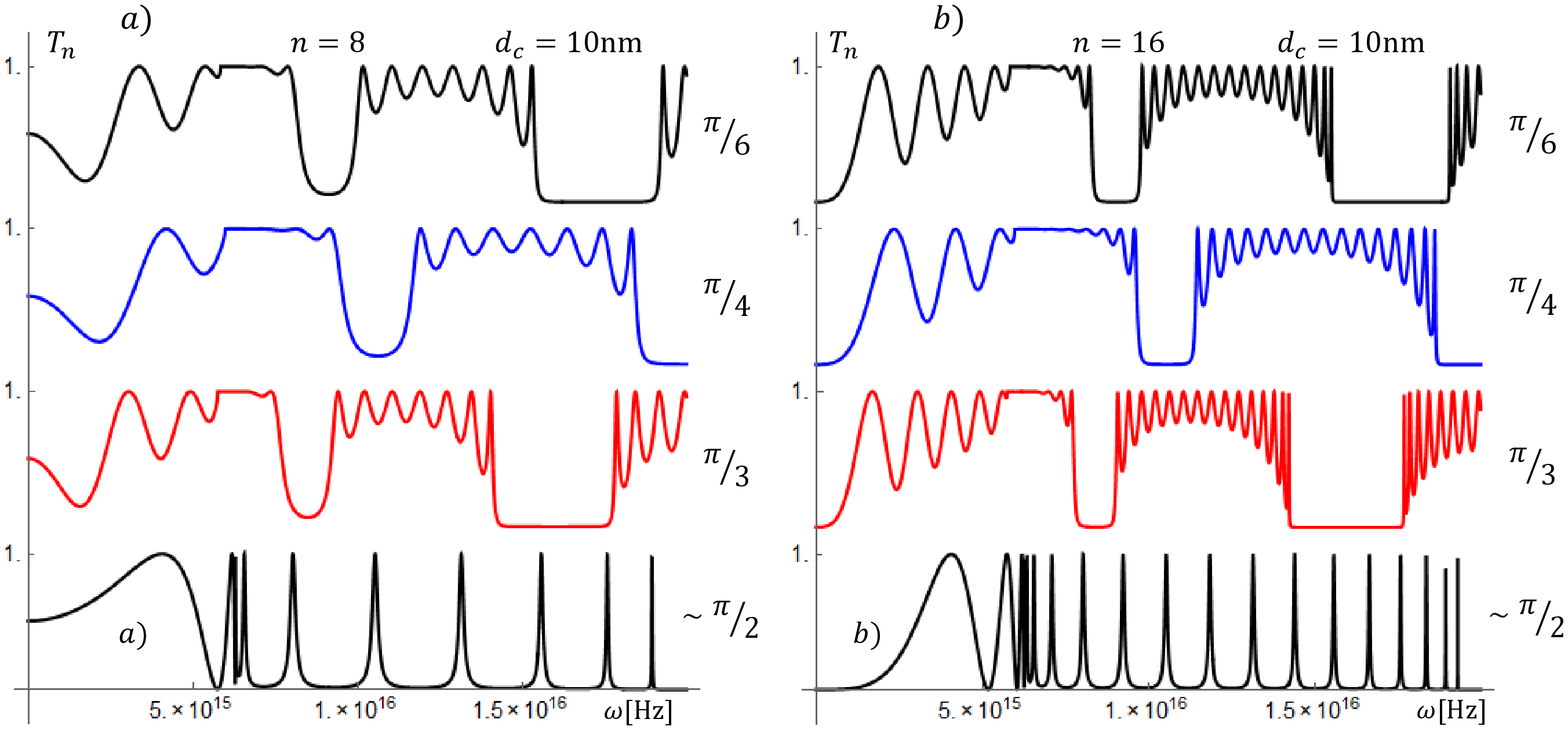}
\caption{Transmittance as a function of the frequency through the metallic superlattices $(air/silver/air)^n$. In a) and b), the thickness of the silver and air slabs is equal to $d_a=$100nm and $d_c=$10nm. In a), the transmittance is shown for $n=8$ and incidence angles $\theta_i=\pi/6,$ $\pi/4$, $\pi/3$, and slightly less than $\pi/2$. Below the plasma frequency $\omega_p=$ 5.75 10$^{15}$rad/s, the transmittance is an oscillating function of $\omega$. Near the incidence angle of $\pi/2$, the transmittance is highly resonant. In b), we repeat the transmittances for $\theta_i=\pi/6,$ $\pi/4$, but now for $n=16$, and we plot also the Kramer condition (black curves) and the dispersion relation of Equation (\ref{RDR}) derived in the theory of finite periodic systems (TFPS) (red lines). It is clear that this recurrence relation predicts the bands and the frequencies of all the resonant plasmons. }\label{fig4}
\end{center}
\end{figure*}

In Figure \ref{fig3}, we show again the transmittance $T_c$ for a single slab, with thicknesses of 80nm (graphs in the left-hand side column) and 1000nm (graphs in the right-hand side column), but now as functions of the incidence angle $\theta_i$, for three values of the EM wave frequency $\omega$. The strong influence of the incoming incidence angle is clear from these graphs. At lower frequencies, the transmittance vanishes unless the incidence is close to the normal incidence. At larger frequencies, it is possible to transmit the EM wave for almost any incidence angle. In the lower panels, the frequency is $\omega = 6.0 x10^{16}$rad/s (corresponding to a wavelength $\lambda \simeq 30nm$), and the transmission is highly oscillating. The number of oscillations is of the order of $d_c/\lambda$. This behavior has also been seen in quantum wells (for energies above the confining potential) where the number of oscillations is of the order of $a/\lambda_B$, $a$ being the well width and $\lambda_B$ the de Broglie wavelength.

In Figure \ref{fig4}, we have the transmittance of photonic superlattices as functions of the frequency for different incidence angles and for different numbers of unit cells. As was frequently stated in the literature of PCs, the analogy with the band structure in electronic superlattices is clear. In photonic superlattices, the role of the incident angle is similar to that of the propagating modes.\cite{Pereyra2015} As is well known from electronic superlattices and from previous calculations for photonic crystals, the position of the band gaps does not change when the number of unit cells increases. Increasing the number of unit cells, the forbidden and allowed bands are better defined, and the reflectance in the forbidden bands is complete. All the properties that can be seen in the behavior of the transmittance of these photonic superlattices are also found in the photonic crystals. It is worth noticing, however, that it is not only much more easy to produce planar structures, it is also much more easy to calculate transfer matrices than the scattering matrices.
\begin{figure*}
\begin{center}
\includegraphics[width=36pc]{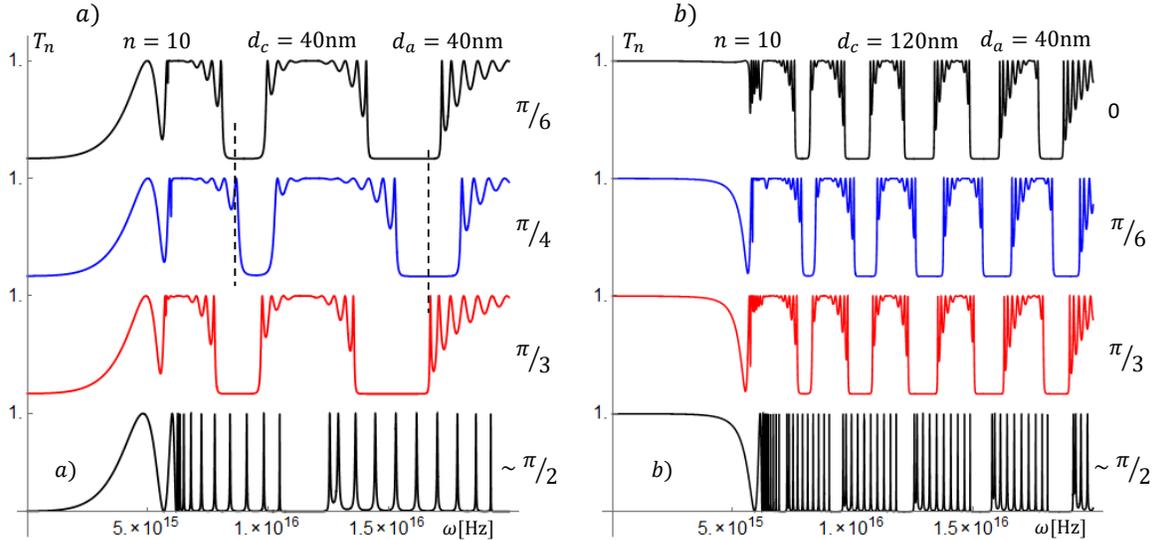}
\caption{Transmittance as a function of the frequency through the metallic superlattices $(air/silver/air)^n$ where the silver and dielectric widths are a) small and equal and b) different with the silver slabs' width being larger. The transmittances are shown for different incidence angles $\theta_i$ indicated on the graphs. Below the plasma frequency $\omega_p=$ 5.75 10$^{15}$rad/s, for small widths, the transmittance is independent of the incidence angle, while for larger silver width, the metallic superlattice is almost completely transparent. For frequencies above $\omega_p$, we have wider bands for small layers' widths and thinner bands for larger silver widths. }\label{fig5}
\end{center}
\end{figure*}
The transmittances shown in Figure \ref{fig4} exhibit other characteristics of the photonic superlattices as a function of the frequency and incidence angle. At low frequencies, i.e., for $\omega < \omega_p$, the transmittance becomes an oscillating function. For $\omega >\omega_p$ and $\theta_i<\pi/2$, we have resonant bands, with intraband plasmons' resonances. Near the incidence angle of $\pi/2$, the transmittance remains highly resonant and the band gaps wider.

It is common to understand the resonant plasmon structure in terms of the SL band structure. In the TFPS, as in other fields of physics like nuclear physics, there is a conceptual distinction between the resonant states through open systems and the energy eigenvalues' spectrum. According to the transmission coefficient in (\ref{ChebRes}), the resonances in the transmission coefficient occur at the zeros of the Chebyshev polynomials $U_{n-1}$. It was shown also in \cite{Pereyra2005} that the energy eigenvalues and the resonant states practically coincide. In the right side column of Figure \ref{fig4}, we plot the transmission coefficients for two cases of the left side column, those for which the incidence angles are $ \theta_i = \pi/4 $ and $ \theta_i = \pi/6 $; this time for $n =$16. As expected, we have more resonances and better defined gaps. The purpose of this column is to plot the Kramer condition (black curves) and the resonant dispersion relation of the TFPS (red lines). Indeed, as we already know, plotting the Kramer condition (see the black curves), we have the allowed and forbidden bands, similar to that of the infinite systems; however, when plotting the derived dispersion relation of the TFPS (see the red lines), we have the bandwidths, as well as the frequencies of all the resonant plasmons. An attempt to determine the intra-sub-band frequencies, based on the dispersion relation of the infinite structures, was done in \cite{Zhang2015}. Therefore, we can state that the resonant band structure is behind the resonant structure of the plasmon modes. In confined systems, the plasmonic modes will be described by the band structure of the energy eigenvalues (see \cite{Pereyra2018}).

To control and manipulate the propagation of electromagnetic fields, one can also play almost at will with the superlattice parameters. In Figure \ref{fig5}, we show two examples where the number of unit cells and the dielectric thicknesses are the same, $n=10$ and $d_a=40$nm, but the silver slabs' thicknesses are in one case equal to $d_a$ and in the other case $d_c=120$nm. When the slabs' thicknesses are both small and equal, we have wider forbidden and allowed bands, and the behavior transmittance below the plasma frequency $\omega_p$ is almost independent of the incidence angle $\theta_i$. However, when the conductor thickness is larger, the low frequency domain becomes transparent, and the forbidden and allowed bands are thinner, as shown in the graph on the right-hand side of Figure \ref{fig5}, where $d_c=$120nm. Many other properties can easily be explored. Among the many applications, the photonic gaps are used as frequency filters. In the left-hand side graph of Figure \ref{fig5}, it is clear (see the dotted lines) that for some frequencies, the EM waves are transmitted for certain incidence angles, while for others, they are blocked. In this graph, we see also that in normal incidence ($\theta_i=0$), the transmittance is complete below the plasma frequency. These results for frequencies below the plasma frequency are strange and occur only when the number of unit cells is larger than one. When the frequencies are larger than $\omega_p$, the band structure remains almost invariant when $n$ varies. It is well known that, when $n$ grows or diminishes, the number of resonances is according to $n$, but the bandwidth remains constant. In general, for open systems, the number of resonances is equal to $n-1$. It is also well known that when $n$ grows, the gaps are better defined. However, when the frequencies are less than $ \omega_p $ and the number of unit cells is greater than one, the coupling of the plasmon modes at low frequencies results in a peculiar oscillating behavior of the transmission coefficient, quite different from a band structure for small frequencies. These oscillations depend largely on $n$ and the width of the conductive slab $d_c$, as will be seen below, related to Figure \ref {fig6}.

There are some analogies and important differences between the electronic transmission coefficients in semiconductor superlattices and the photonic transmittance in metallic superlattices. The analogue of the barrier height in the electronic SLs is the plasma frequency $\omega_p$. While the transmission coefficient in electronics SLs has isolated resonances or mini-bands below the threshold, in the metallic SLs, the photonic transmittance becomes an oscillating function of $\omega$. In the electronic SLs, the widths of the bands grow with the energy, and soon, they overlap. In the metallic SLs, the width of the allowed bands remains almost constant; the same happens with the width of the forbidden gaps.

\begin{figure*}
\begin{center}
\includegraphics[width=34pc]{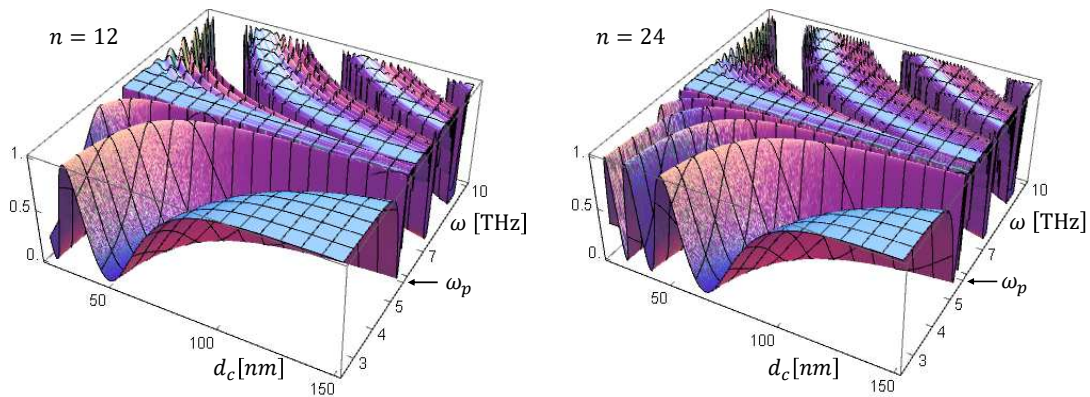}
\caption{Transmittance through $(air/silver/air)^n$ SLs
 as a function of the frequency $\omega$ and the conductor-layer width $d_c$. In a) and c), the number of unit cells $n$ is odd, while in b) and d), it is even. The difference in the behavior of $T_n$ below and above $\omega_p$ is clear in the upper panels. The line shapes of resonant plasmons with $\omega>\omega_p$ are thinner, therefore more localized, than the plasmon frequencies at low frequencies. We assume that the latter result from a complex coupling of interface plasmons. A parity effect is also clear for $d_c$ larger than $\sim$50nm. At larger conductor widths, the low frequency transmission either vanishes, for $n$ odd, or tends asymptotically to one, for $n$ even. In the lower-panel graphs, we plot, together with the transmission coefficient, the dispersion relation of Equation (\ref{RDR}), and we see that, if $n$ is odd, the number of plasmon resonances is the same for SLs with $n$ and with $n+1$ unit cells. }\label{fig6}
\end{center}
\end{figure*}

Here, we also see that the photonic transmittance through the metallic superlattice has different characteristics in the two frequency domains limited by the plasma frequency $\omega_p$. Our results also exhibit important differences and some unusual features. In Figure \ref{fig6}, we show the transmission coefficients in 3D graphs (upper panels) and, in the lower panels, for a fixed $\omega$, smaller than $\omega_p$. In the upper panels, we assume a constant dielectric width $d_a=40$nm, and for the lower panels, the dielectric width is $d_a=20$nm and the incoming wave frequency $\omega=1.8849 10^{15}$rad/s. The incidence angle in the upper panels is $\pi/6$ and in lower panels is $\pi/3$. The line shapes of the resonances for $\omega>\omega_p$ are, in general, thin and thinner when the number of unit cells grows; this characteristic shows that these plasmons are highly localized, as one expects from surface plasmon polaritons. The transmission for frequencies $\omega<\omega_p$ is, as mentioned before, an oscillating behavior, with shorter tunneling times. These fast plasmons result possibly from the coherent coupling of interface plasmons. In the upper panels of Figure \ref{fig6}, we see also a feature that was observed in Figure \ref{fig5}: the decreasing of the bandwidths when the conductor layer $d_c$ is increased.

An important and challenging property that we recognize in the graphs in Figure \ref{fig6} is an apparent parity effect at larger conductance widths. To visualize this effect, the graphs in the left column correspond to $ n $ odd, while the graphs in the right column correspond to $ n $ even. Growing the conductor width, with fixed dielectric width $d_a$, we see that either the transmission coefficients tend to zero ($n$ odd) or tend asymptotically to one ($n$ even). It is not yet clear what kind of coupling is behind this effect. The rather complex coherent coupling or superposition of the plasmon modes depends on many parameters, and eventually, the coherent superposition gives rise to the familiar allowed and forbidden bands, which we see for $\omega>\omega_p$. Apparently, the coupling of low frequency plasmons leads to a kind of transparency that deserves further analysis. In \cite{Garbecht2017}, a decrease of the plasmon resonance frequency was observed in metal/semiconductor TiN/(Al,Sc)N multilayers when the interlayer thickness was increased. The authors suggested that this effect
resulted from resonant coupling between bulk and surface plasmons
across the dielectric interlayers.

We showed here a simple and accurate theory to study the transport properties of electromagnetic fields through a metallic superlattice, and we showed that playing with the SL parameters, the whole spectrum of features and effects that characterize the photonic crystals can also be observed for the transmission coefficients. When $\omega > \omega_p$, the transmittance is characterized by a resonant band structure with wide or thin bandwidths. We showed that the position of the bands is highly sensitive to the incidence angle $\theta_i$; the number of resonances in the bands is determined by the number of unit cells $n$. As $n$ grows, the reflectance in the stopbands is complete. When $\omega < \omega_p$, we predict interesting oscillations, as well as attenuation or amplification effects of the transmission coefficients.

\section{Transmittance of EM Waves through Left-Handed Photonic Superlattices}\label{section4}

Negative refraction index (left-handed) materials have become the object of an active and controversial research field recently, with striking and new effects. Veselago \cite{Veselago1968} predicted in 1967 some unique properties of EM wave propagation in LHM: (a) the waves appear to propagate towards the source and not away from it; (b) their group velocity is negative; and (c) because waves incident on RH/LH interfaces are refracted to the same side of the normal, converging and diverging lenses exchange their roles. Furthermore, Veselago proposed the constraints:
\begin{eqnarray}
\frac{\partial \epsilon (\omega)}{\partial \omega} > 0 \,\,\,\,\,\,\,\,\,\,\,\,\,\,\,\,\,\,\frac{\partial \mu (\omega)}{\partial \omega} > 0 \label{Veselago}
\end{eqnarray}
for the energy transferred from the source to the load to be positive and to avoid causality violations. Smith and Kroll \cite{Smith2000}, on the other hand, maintained that while a reversed $k$ resembles a time-reversed propagation towards the source, the work done was nevertheless positive. Incidentally, by analyzing the implications of Eq. (\ref{Veselago}), these authors reached the conclusion that the group velocity ought to be positive for both types of materials. A simplistic approach would lead to the conclusion that the sign in $\pm k$ causes opposing group velocities.

Besides the blazing presumption of perfect lenses \cite{Pendry2003} and problems like the sign selection and direction of motion of the energy and the electromagnetic field, it has been explicitly shown that the transmission amplitude of a single left-handed slab is just the complex conjugate of the transmission amplitude of a similar, but right-handed slab. An important consequence of this property is that the phase time $\tau$ of a single slab, defined as the frequency derivative of the transmission-amplitude's phase $\theta_t$, becomes not only the negative of the corresponding phase time of a right-handed slab; it implies in general negative transmission times, which results in warnings of possible causality violation.

The transmission amplitude of a left-handed slab, with refraction index $n_L$, bounded by semi-infinite right-handed media (with refraction index $n_2$), is (see \cite{Pereyra2019}):
\begin{widetext}
\begin{eqnarray}
t_L=\left[\cos {(k_Ld_L\cos \theta_L)}+\frac{i}{2|n_L|n_2}\left(n_2^2\frac{\cos {\theta_L}}{\cos {\theta_2}}+n_L^2\frac{\cos {\theta_2}}{\cos {\theta_L}}\right)\sin {(|k_L|d_L\cos \theta_L)}\right]^{-1},
\end{eqnarray}
while the transmission amplitude of right-handed media, with refraction index $n_R$, bounded also by semi-infinite right-handed media (with refraction index $n_2$), is (see, for example, \cite{Simanjuntak2007}):
\begin{eqnarray}
t_R=\left[\cos {(k_Rd_R\cos \theta_R)}-\frac{i}{2n_Rn_2}\left(n_2^2\frac{\cos {\theta_R}}{\cos {\theta_2}}+n_R^2\frac{\cos {\theta_2}}{\cos {\theta_R}}\right)\sin {(k_Rd_R\cos \theta_R)}\right]^{-1}.
\end{eqnarray}
\end{widetext}
The phase time $\tau_L$ of a left-handed slab is then the negative of the corresponding phase time of a right-handed slab. This property suggests, naturally, the possibility of the violation of the causality principle. Gupta et al. studied also the transmission of electromagnetic pulses across a parallel slab of a medium where $\epsilon$ and $\mu$ are functions of the frequency $\omega$ and found ranges of frequency where the delay time is positive and ranges where it is negative \cite{Gupta2004}. Recently, this problem was studied in detail {\cite{Pereyra2011}}, for negative, but constant $\epsilon$ and $\mu$, and the phase time predictions were shown to describe the actual wave packet evolution, but with clear evidence of an optical antimatter behavior.

The analysis in multilayered structures is much more involved. In the multilayer structures, it is no longer true that the transmission amplitude $t_L$ of an SL with alternating LH and RH layers is the complex conjugate of the transmission amplitude $t_R$ of the corresponding SL where the LH layers are replaced by RH layers with equal, but positive refraction indices.
\begin{figure*}
\begin{center}
\includegraphics[width=26pc]{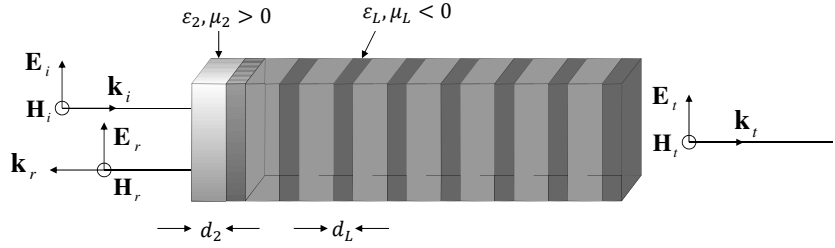}
\caption{A metamaterial superlattice $air(L R_2)^nair$ with $n$ unit cells, where left- and right-handed media alternate. We assume the normal incidence of the electromagnetic field with parallel polarization.}\label{fig7}
\end{center}
\end{figure*}

We will now calculate the transmission coefficient through a metamaterial superlattice $(LR)^n$ with $n$ unit cells bounded by semi-infinite air layers, for the unit cell parameters indicated in Figure \ref{fig7}. The transmission amplitude $t_{aSa}$ through the structure $air(LR)^{n}air$ is given by \cite{Simanjuntak2007}:
 \begin{eqnarray}
 t_{aSa}=\frac{1}{\alpha_{aSa}}.
 \end{eqnarray}
where $\alpha_{aSa}$ is the element $(1,1)$ of the transfer matrix that connects electromagnetic fields at the left- and right-hand sides of the $air(LR)^n air$ structure. Assuming $|\mu_L|=\mu_2\simeq \mu_o$, this matrix element is \cite{Pereyra2000}:
 \begin{eqnarray}
 \alpha_{aSa}=\alpha_{nr}+i\left(\frac{1+n_2^2}{2 n_2}\alpha_{ni}+\frac{n_2^2-1}{2 n_2}\beta_{ni}\right).
 \end{eqnarray}
\begin{widetext}
Here, $\alpha_{nr}=U_n-\alpha_r U_{n-1}$, $\alpha_{ni}=-\alpha_{i}U_{n-1}$,
$\beta_{nr}=\beta_r U_{n-1}$, and $\beta_{ni}=\beta_{i}U_{n-1}$ are the real and imaginary parts of the matrix elements of the $n$ cell (superlattice) transfer matrix $M_n=\left( \begin{array}{ll}\alpha_n & \beta_n\cr \beta_n^*& \alpha_n^* \end{array} \right)$. $\alpha_r$, and $\alpha_i$, $\beta_r$ and $\beta_i$ are the real and imaginary parts of the matrix elements:
\begin{eqnarray}\label{MMSLalf}
 \alpha=e^{ik_2d_2\cos \theta_2}\left(\cos {(k_Ld_L\cos \theta_L)}+\frac{i}{2n_1n_2}\left(n_2^2\frac{\cos {\theta_L}}{\cos {\theta_2}}+n_L^2\frac{\cos {\theta_2}}{\cos {\theta_L}}\right)\sin {(k_Ld_L\cos \theta_L)}\right),
 \end{eqnarray}
 and:
\begin{eqnarray}
 \beta=\frac{ie^{ik_2d_2\cos \theta_2}}{2n_Ln_2}\left(n_L^2\frac{\cos {\theta_2}}{\cos {\theta_L}}-n_2^2\frac{\cos {\theta_L}}{\cos {\theta_2}}\right)\sin {(k_Ld_L\cos \theta_L)},
 \end{eqnarray}
\end{widetext}
of the single-cell transfer matrix $M$. As mentioned before, $U_n$ is the Chebyshev polynomial of the second kind and order $n$, evaluated at the real part of $\alpha$.
\begin{figure}
\begin{center}
\includegraphics[width=20pc]{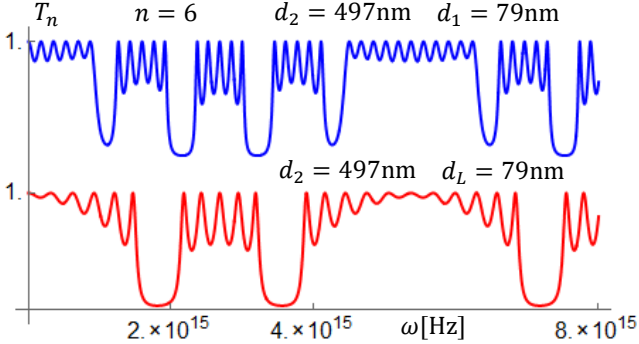}
\caption{The transmission coefficients for two superlattices $air(RR)^n air$ (upper panel) and $air(LR)^n air$ (lower panel), with the same parameters except for the signs of the refraction indices $\epsilon_1$ and $\mu_1$. For the upper panel, we have $n$=6, $n_1=2.22$, $n_2=1.41$, $d_1=79$nm, and $d_2=497$nm. For the lower panel, we have $n$=6, $n_1=-2.22$, $n_2=1.41$, $d_1=79$nm, and $d_2=497$nm.}\label{fig8}
\end{center}
\end{figure}

In Figure \ref{fig8}, we have the transmission coefficients for two superlattices $air(RR)^n air$ and $air(LR)^n air$, with the same parameters except for the signs of the refraction indices $\epsilon_1$ and $\mu_1$. For the upper panel, we have $n$=6, $n_1=2.22$, $n_2=1.41$, $d_1=79$nm, $d_2=497$nm. For the lower panel, we have $n$=6, $n_1=-2.22$, $n_2=1.41$ and $d_1=79$nm, and $d_2=497$nm. As for the metallic superlattice, the transmittance has a band structure; however, at variance with the metallic superlattices, there is no threshold frequency.

\begin{figure*}
\begin{center}
\includegraphics[width=20pc]{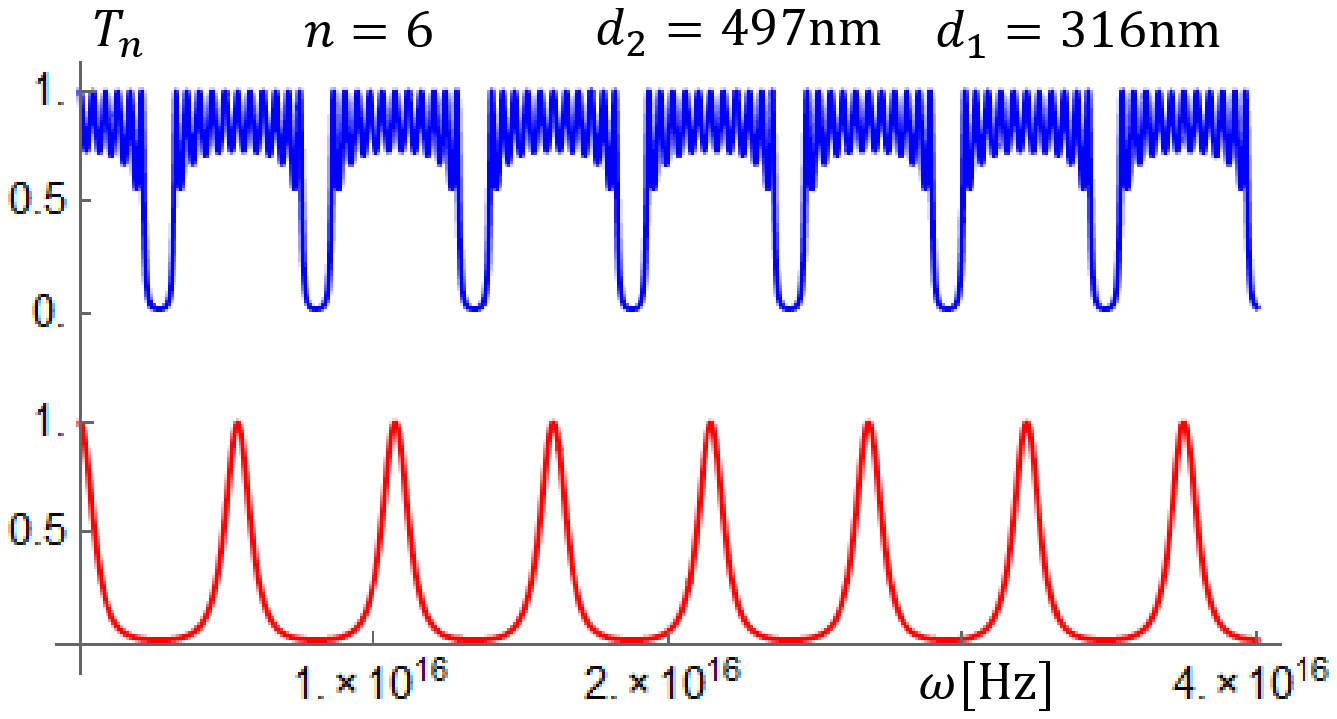}
\caption{Transmission coefficients (TCs) for the superlattices $air(RR)^6 air$ and $air(L R_2)^6air$, as functions of the frequency $\omega$, when $d_L=d_1=d_2 n_2/n_L=316.5$nm and $d_2 =497$nm. The TCs of the superlattices $air(LR)^6 air$ become a sequence of isolated and equidistant peaks, while the TCs of the superlattices $air(RR)^6 air$ become a periodic sequence of resonant bands. This characteristic is independent of the number of unit cells.}\label{fig9}
\end{center}
\end{figure*}

In the left-handed superlattices, an interesting property appears when the well-known $\lambda/4$ relation for the layers' widths is considered. If we choose the widths such that $d_L=d_1=d_2 n_2/n_L=316.196$nm, the transmission coefficients become as shown in Figure \ref{fig9}. In the upper panel for the superlattice $air(RR)^n air$, we have a periodic band structure, while for the superlattice $air(LR)^n air$, in the lower panel, the bands collapse into a periodic sequence of isolated peaks.

\begin{figure*}
\begin{center}
\includegraphics[width=34pc]{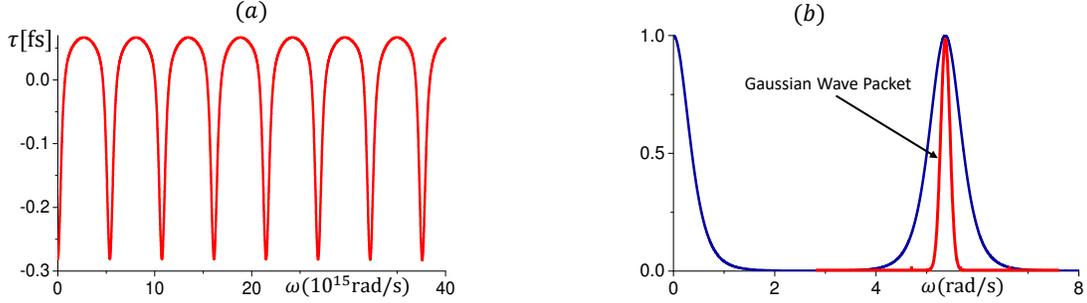}
\caption{In Panel (a), the tunneling (phase) time through the structure $air(LR)^7air$, whose transmission coefficient is shown in the lower panel of Figure \ref{fig9}. The tunneling times of the wave packet (WP) peaks through the $air(LR)^7air$ structure, obtained from $\tau=\partial t_{aSa}/\partial \omega $, at the resonant frequencies, are $\cong -0.28$fs. In b), a Gaussian wave packet with the centroid at the resonance with frequency $\omega \simeq 5.32 \times 10^15$rad/s, and the envelope as shown here is prepared, at $t=0$, at a distance $z_0=L=40l_c$ from the SL.}\label{fig10}
\end{center}
\end{figure*}

\begin{figure*}
\begin{center}
\includegraphics[width=21pc]{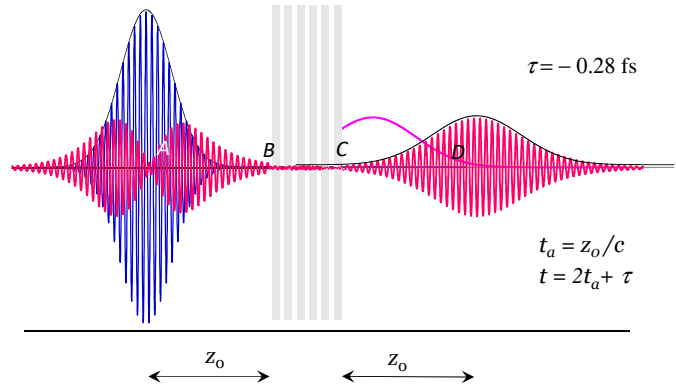}
\caption{A Gaussian WP with the centroid at the resonance with frequency $\omega \simeq 5.32 \times 10^{15}$rad/s is prepared, at $t=0$, at a distance $z_0=L=40l_c$ from the SL (see the blue curve). The WP moves towards $D$, passing through the SL $(LR)^6$, and reaches the point $D$ (see the red curves), as predicted, at $t=\tau_D=2z_0/v_g+\tau \simeq 54$.$02$fs. The WP is partially transmitted and partially reflected. Not all frequencies are transmitted because the transmission coefficients at the center of the packet are larger than in the tails. Thus, a dip is formed in the reflected (red) WP. The distance between the peak of the (blue) WP and the dip in the reflected (red) WP depends on $\tau$.}\label{fig11}
\end{center}
\end{figure*}

In this particular relation of the metamaterial superlattices, when $d_L=d_1=d_2 n_2/n_L$, also the phase time becomes a periodic function of $\omega$ and negative around the resonant frequencies. As can be seen in the graph at the left side of Figure \ref{fig11}, the phase times at the resonant frequencies of the transmission coefficient of the SL $air(LR)^6 air$ are negative. If we build a wave packet moving through an SL $(LR)^7$, with the centroid at the peak of one of the resonant transmission peaks (see the Gaussian envelope in Figure \ref{fig10}), the predicted tunneling time of the wave packet (WP) peak through this SL with length $L=6(d_L+d_2)=6 l_c$, is $\tau=-0.187396$fs, and the space-time evolution is as shown in Figure \ref{fig11}.

A theoretical simulation shows that the centroid of the WP, at $t=0$, is at $-z_0\simeq L$ (the blue curve in Figure \ref{fig11}). Moving with group velocity $v_g=c$, the WP arrives at point $B$ at $t_B=z_0/v_g\simeq 27$.$154$fs. If the WP were to move inside the SL with the same velocity $v_g$, its peak would reach $C$ at $t_C=(z_0+L)/v_g\simeq 27$.$569$fs and the point $D$ at $t_D=(2 z_0+L)/v_g\simeq 54$.$72$ fs. However, that does not happen. Plotting the space-time evolution of the WP, we see that it reaches the point $D$, as predicted, at $\tau_D=2z_0/v_g+\tau \simeq 54$.$02$fs. The red WPs are the partially transmitted and partially reflected WPs. What happens? Do we have an antimatter effect? This result confirms only the possibility of negative tunneling times. The features of the reflected and transmitted wave packets depend on the frequencies of the wave packet components. The components of the wave packet for this simulation were defined near the resonant peak; see Figure \ref{fig10}. These components have, on the one hand, larger transmission coefficients and, on the other, negative tunneling times. This is the reason why the WP components near the centroid of the incoming WP are transmitted while those in the tails, the frequencies of which differ from the centroid frequency, get reflected with, apparently, two new wave packets.

In these systems, as in the metallic superlattices, the dispersion relation is also behind the resonant band structure. Indeed, in Appendix B, we consider other MMSLs with special relations between the layers' widths and show, for each of these cases, the close relation between the resonant band structure of the transmission coefficient and the dispersion relation in Equation (\ref{RDR}). As shown in Appendix B, both the Kramer condition and the resonant dispersion relation predict that the bandwidth reduces to one point, at those frequencies determined by the equation:
\begin{equation}
 1=\alpha_r
\end{equation}
with $\alpha_r$ the real part of the matrix element given in (\ref{MMSLalf}).
The phase time behavior for MMSLs depends on many factors. Different Gaussian packets with centroids at qualitatively different frequency domains will have also different space-time evolutions.

\section{Conclusion}
The transmittance of electromagnetic waves through metallic and left-handed superlattices was calculated, exhibiting all the physical properties of the widely studied photonic crystals. These results showed that not only much simpler devices, but also easier theoretical calculations were possible. A careful analysis of the space-time evolution showed that there was no violation of the causality principle.

\vspace{6pt}

\appendix
\section{Resonances and Resonant Dispersion Relation }
In [23] and [35], the relation $|\alpha_n|^2=1+|\beta|^2$ was replaced into:
\begin{equation}
T_n=\frac{1}{|\alpha_n|^2}
\end{equation}
to write the transmission coefficient in the form:
\begin{equation}
T_n=\frac{1}{1+ |\beta_n|^2},
\end{equation}
which, recalling the relation $\beta_n=\beta U_{n-1}$, is written as:
\begin{equation}
T_n=\frac{1}{1+ |\beta|^2 |U_{n-1}(\alpha_R)|^2}.
\end{equation}

It is clear that in this representation, the transmission coefficient becomes one when the energy is such that $\alpha_R$ becomes a zero of the Chebyshev polynomial $U_{n-1}$. It is well known that the zeros of $U_{n-1}$ are at the points defined by $\cos{\nu \pi/n}$, with $\nu$ = 1,2,...,$n$ 1. This relation was generalized in \cite{Pereyra2017} to include the quantum number $\mu$ that labels the bands and the quantum number $\nu$ that labels the intra-band energy levels. In terms of these quantum numbers, the dispersion relation that defines the bands and the intra-band levels is:
\begin{widetext}
\begin{equation}
\cos \frac{\nu+(\mu-1)n}{n}\pi=(\alpha_R)_{\mu,\nu}\hspace{0.3in} {\rm with} \hspace{0.3in}\mu=1,2,3,... \;\; \nu=1,2,...,n-1.
\end{equation}
\end{widetext}
Because of the close relation with the resonant behavior in open systems, it will be referred to as the resonant dispersion relation (RDR). Another important relation that was derived in the theory of finite periodic systems (see \cite{Pereyra1998,Pereyra2002}) is the Landauer conductance $G_n$, which in the particular case of only one propagating mode becomes:
\begin{eqnarray}
G_n=G\frac{1}{U_{n-1}^2}
\end{eqnarray}

Here, $G$ is the conductance of a unit cell. This result is very insightful, because $G_n$ is written in terms of two factors, each of which has a close relation with a fundamental quantum property, the quantum tunneling and the quantum coherence, represented by $G$ and $1/U_{n-1}^2$, respectively. In the theory of finite periodic systems, the polynomial
$p_{N,n}$ in the multi-mode case or the Chebyshev polynomials $U_{n}$ in the one propagating mode limit carry the physical information about the number of layers, $n$, the number of channels, $N$, and more importantly, the information of the complex interference phenomena, responsible for the band structure.

\section{The Quarter-Lambda Band Structure and the RDR}

\begin{figure*}
\begin{center}
\includegraphics[width=36pc]{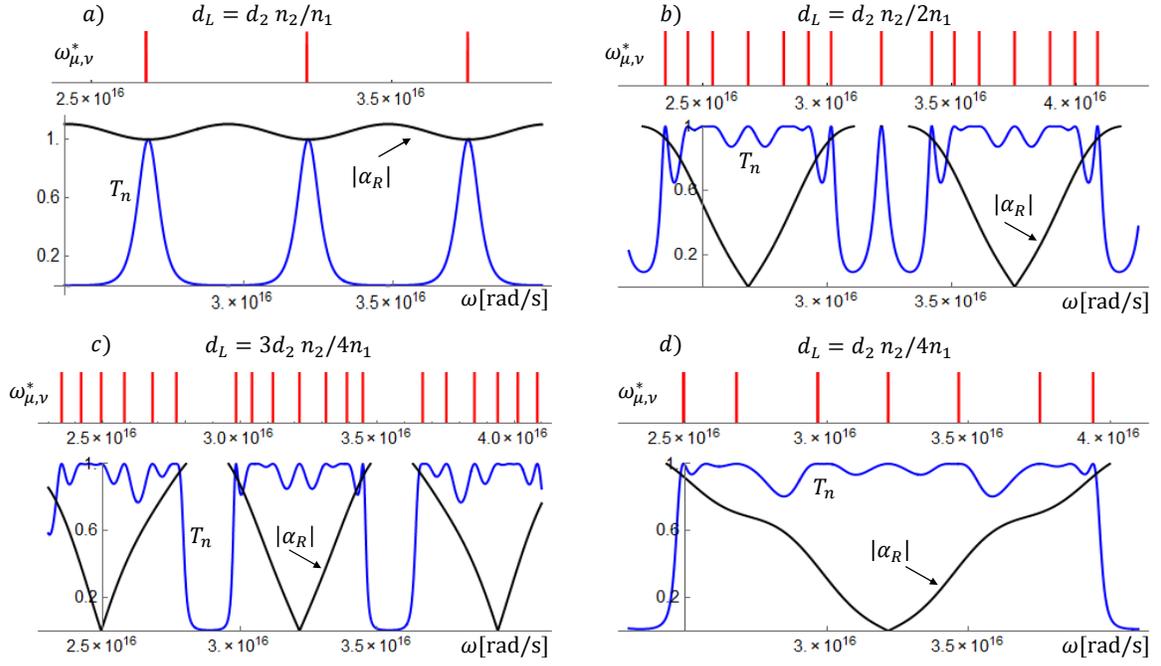}
\caption{ In this figure, we plot the transmission coefficient for the quarter lambda in the manuscript and for other relations, indicated in the graphs. The resonant transmission coefficient (blue curves) are plotted together with the dispersion relations for an infinite periodic structure (black curves) and with the dispersion relation of the theory of finite periodic systems (red lines). In (a), the narrow bands in the quarter lambda limit agree with the predictions of the Kramer condition and with those of the resonant dispersion relation. In the other panels, we find again that the resonant dispersion relation predicts the resonances of the transmission coefficients.}\label{QL}
\end{center}
\end{figure*}

In Figure \ref{QL}, we repeat the transmission coefficient in the lower panel of figure, and we plot the transmission coefficient for the other cases, for the parameters indicated in the panels. The close relation between the resonant behavior between the resonant behavior and the dispersion relations, in the special case and others, is indicated in the panels. In the special case of Figure \ref{QL}a, the allowed energies are those where $|\alpha_R|=1$ (see the black curve in Figure \ref{QL} a). A similar reduction occurs in the resonant dispersion relation of Figure \ref{QL}b around $\omega=3.2 10^{16}$Hz. If we denote the dispersion relation angle as:
\begin{eqnarray}
 \theta_{\mu,\nu}^{(n)}&=& \frac{\nu+(\mu-1)n}{n}\pi
\end{eqnarray}
The narrow bands in Figure \ref{QL}a (reduced to isolated resonances for $d_L=d_2 n_2/n_1$) occur when $\nu=n$ and $\mu$ is such that:
\begin{eqnarray}
 \theta_{\mu,\nu}^{(n)}&\rightarrow & s 2 \pi \hspace{0.3in}{\rm with}\hspace{0.3in} s=1,2,3,...
\end{eqnarray}
Similarly, when $d_l=d_2 n_2/2n_1$, some bands become narrow bands (see Figure \ref{QL}b). The resonances corresponding to these narrow bands occur when $\nu=n$ and $\mu$ such that:
\begin{eqnarray}
 \theta_{\mu,\nu}^{(n)}&\rightarrow & s \pi \hspace{0.3in}{\rm with}\hspace{0.3in} s=1,2,3,...
\end{eqnarray}
\end{document}